\documentclass{vgtc}                          

\ifpdf
  \pdfoutput=1\relax                   
  \usepackage{graphicx}                
  \DeclareGraphicsExtensions{.pdf,.png,.jpg,.jpeg} 
\else
  \usepackage{graphicx}                
  \DeclareGraphicsExtensions{.eps}     
\fi%

\graphicspath{{figures/}{pictures/}{images/}{./}} 

\usepackage{microtype}                 
\PassOptionsToPackage{warn}{textcomp}  
\usepackage{textcomp}                  
\usepackage{mathptmx}                  
\usepackage{times}                     
\usepackage{cite}                      
\usepackage{tabu}                      
\usepackage{booktabs}                  
\usepackage[colorlinks = true,
            linkcolor = blue,
            urlcolor  = blue,
            citecolor = blue,
            anchorcolor = blue]{hyperref}
\newcommand{\MYhref}[3][blue]{\href{#2}{\color{#1}{\underline{#3}}}}%

\usepackage{dblfloatfix}
\usepackage{enumitem}
\setlist[itemize]{noitemsep, topsep=0pt}

\newcommand{\para}[1]{\vspace{1mm}\noindent\textbf{#1}}
\newcommand{\cmo}{\textcolor[rgb]{0, 0, 0}}

\newsavebox{\coloredquotationbox}
\newenvironment{coloredquotation}
 {%
  \begin{trivlist}
  \begin{lrbox}{\coloredquotationbox}
  \begin{minipage}{\dimexpr\linewidth-2\fboxsep}
 }
 {%
  \end{minipage}
  \end{lrbox}
  \item\relax
  \parbox{\linewidth}{
    \begingroup
    \color[RGB]{224,215,188}%
    \hrule
    \color[RGB]{249,245,233}%
    \hrule
    \color[RGB]{224,215,188}%
    \hrule
    \endgroup
    \colorbox[RGB]{249,245,233}{\usebox{\coloredquotationbox}}\par\nointerlineskip
    \begingroup
    \color[RGB]{224,215,188}%
    \hrule
    \color[RGB]{249,245,233}%
    \hrule
    \color[RGB]{224,215,188}%
    \hrule
    \endgroup
  }
  \end{trivlist}
 }

\onlineid{0}

\vgtccategory{Research}

\vgtcinsertpkg

\title{Beyond Generating Code: Evaluating GPT on a Data Visualization Course}

\author{Chen Zhu-Tian\thanks{John A. Paulson School Of Engineering And Applied Sciences, Harvard University. E-mail: \{ztchen, jtroidl, simonwarchol, jbeyer, pfister\}@seas.harvard.edu} %
\and Chenyang Zhang\thanks{Research intern in Harvard. E-mail: zhang414@illinois.edu} %
\and Qianwen Wang\thanks{Biomedical Informatics, Harvard Medical School. E-mail:\{qianwen\_wang, nils\}@hms.harvard.edu }
\and Jakob Troidl\footnotemark[1]
\and Simon Warchol\footnotemark[1]
\and Johanna Beyer\footnotemark[1]
\and Nils Gehlenborg\footnotemark[2]
\and Hanspeter Pfister\footnotemark[1]
}

\teaser{
  \centering
  \includegraphics[width=\linewidth]{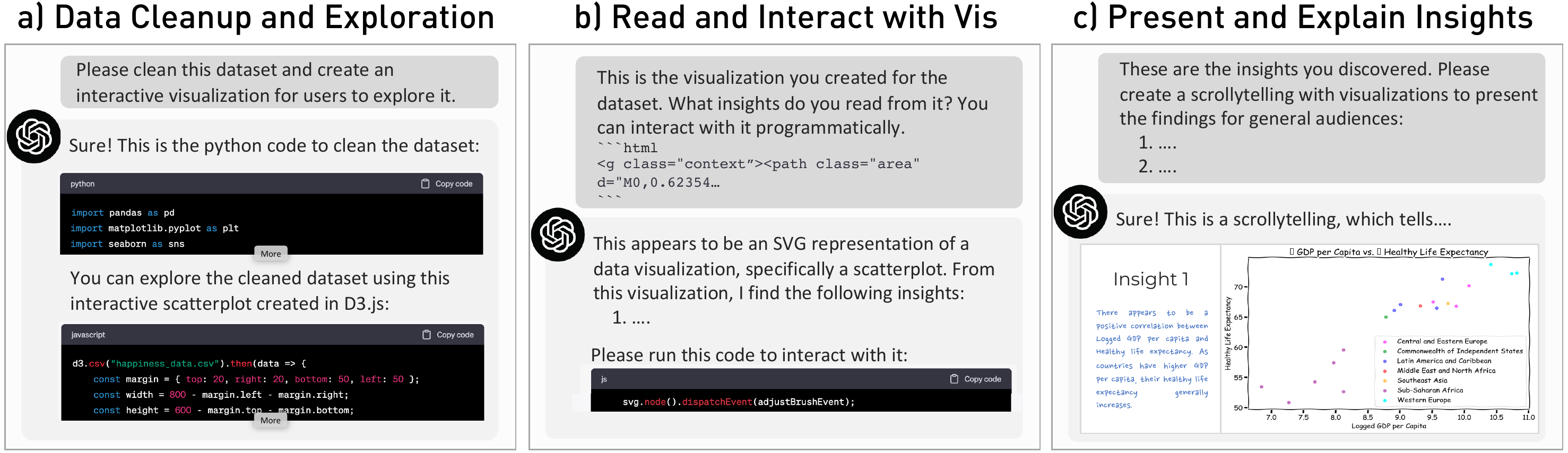}
  \caption{Our experiments show that GPT can a) clean and explore CSV datasets,
  b) read visualizations in SVG format,
  interact with visualizations through dispatching Javascript events,
  and c) create explanatory visualizations to present data insights.
  }
  \label{fig:teaser}
}

\abstract{This paper presents an empirical evaluation of the performance of the Generative Pre-trained Transformer (GPT) model in Harvard's CS171 data visualization course.
While previous studies have focused on GPT's ability to generate code for visualizations, this study goes beyond code generation to evaluate GPT's abilities in various visualization tasks, such as data interpretation, visualization design, 
visual data exploration, and insight communication.
The evaluation utilized GPT-3.5 and GPT-4 \cmo{through the APIs of OpenAI} to complete assignments of CS171,
and included 
a quantitative assessment based on the established course rubrics,
a qualitative analysis informed by the feedback of three experienced graders,
and an exploratory study of GPT's capabilities in completing border visualization tasks.
Findings show that GPT-4 scored 80\% on quizzes and homework, 
and \cmo{Teaching Fellows} could distinguish between GPT- and human-generated homework with 70\% accuracy.
The study also demonstrates GPT's potential in completing various visualization tasks, such as data cleanup, interaction with visualizations, and insight communication. 
The paper concludes by discussing the strengths and limitations of GPT in data visualization, 
potential avenues for incorporating GPT in broader visualization tasks, 
and the need to redesign visualization education.
} 

\begin{document}


\firstsection{Introduction}

\maketitle

\cmo{Recent years have witnessed remarkable advances in Natural Language Processing (NLP), 
especially with models like the Generative Pre-trained Transformer (GPT).}
GPT's ability to generate coherent and contextually appropriate text has been widely explored across various domains, including the field of data visualization.
For example, LIDA uses GPT-3 to generate grammar-agnostic visualization specifications for a given dataset~\cite{Dibia2023LIDA}.

\cmo{As we delve deeper into this nexus of NLP and data visualization, the pressing question emerges: 
\textbf{\emph{How proficient is GPT in data visualization tasks?}}
While its text-generating capabilities are well-documented, the depths of GPT's abilities to interpret, design, explore, and communicate visual data remain uncharted.}
Previous studies have explored and evaluated GPT's abilities on visualization tasks but mainly focus on generating code for creating data visualizations~\cite{Dibia2023LIDA, maddigan2023chat2vis}.
Yet, applying machine learning (ML) methods in data visualization goes beyond mere code generation. 
Rather, ML methods can be applied to various visualization tasks,
including design principles, data interpretation, visual encoding, user understanding, and insight communication~\cite{wang2021ml4vis}.

\cmo{To address our central research question regarding GPT's aptitude in data visualization, 
we drew inspiration from methodologies in other domains.}
Standardized exams can effectively assess GPT's proficiency in various tasks within a specific domain, 
such as medicine~\cite{nori2023capabilities} and law~\cite{choi2023chatgpt}.
Although there is no established standardized examination for data visualization, 
the assignments in a data visualization course can serve as a means to evaluate GPT's capabilities in various visualization skills.
\cmo{In light of this gap, 
we innovatively \textbf{\emph{employed the assignments of a data visualization course as our evaluative tool}}. 
This approach offered us a comprehensive and rigorous framework to critically assess GPT's capabilities across a spectrum of visualization tasks.}

This paper presents a set of evaluations of GPT's performance in the context of a data visualization course: Harvard's CS171\footnote{https://www.cs171.org. \cmo{Accessed 08.10.2023.}}.
Launched in 2008, CS171 is an introductory data visualization course that has engaged thousands of students and evolved through iterative improvements in the past 16 years. 
Specifically, we utilized GPT-3.5 and GPT-4 (without any specialized prompt crafting or interactions) \cmo{through the APIs of OpenAI} to complete course assignments, i.e., a set of challenge problems developed to evaluate students' mastery of various visualization skills.
These assignments were then randomly mixed with past student submissions and graded by three former Harvard CS171 Teaching Fellows (TFs). 
We present 1) a quantitative assessment based on the established course rubrics, 2) a qualitative analysis informed by the feedback of three experienced graders,
and 3) an exploratory study of GPT's capabilities in completing visualizations projects.
We find that GPT-4 could score 80\% on the quizzes and homework of CS171,
and the TFs were able to distinguish between GPT- and human-generated homework with 70\% accuracy based on certain heuristics.
The exploratory study implied that GPT could complete various visualization tasks,
such as data cleanup and exploration (\autoref{fig:teaser}a),
reading and interacting with visualizations (\autoref{fig:teaser}b),
and presenting and explaining insights (\autoref{fig:teaser}c).
Our evaluations provide valuable insights into GPT's strengths and limitations in the context of data visualization, 
suggesting potential avenues for incorporating GPT into broader visualization tasks,
and prompting a reconsideration of how we approach visualization education.
All supplemental materials can be found at: \url{https://github.com/GPT4VIS/GPT-4-CS171}

\section{Related Work}

Large language models (LLMs) refer to neural network-based models that are trained on massive amounts of data to generate human-like language responses. 
One of the most well-known examples of such models is GPT 
(e.g., GPT-3~\cite{brown2020language}, GPT-4~\cite{bubeck2023sparks}, ChatGPT~\cite{chat-gpt}), developed by OpenAI.
In recent years, a growing body of research has delved into GPT's potential, showcasing remarkable performance across various NLP tasks, including generation, understanding, reasoning, and translation~\cite{bubeck2023sparks, floridi2020gpt, wei2022chain, dale2021gpt}.
Recent studies show that GPT-3 and GPT-4 can pass a range of professional exams,
including the Law School Admission Test (LSAT)~\cite{choi2023chatgpt}, 
the bar exam~\cite{katz2023gptbar},
the Graduate Record Examination (GRE)~\cite{gpt-4}, 
and the USMLE~\cite{nori2023capabilities}, among others.
Some researchers even argue that GPT-4 could be seen as an early version of artificial general intelligence (AGI)~\cite{bubeck2023sparks}. 

In the field of visualization, 
there have also been many works exploring the use of LLMs, but most of them have focused on generating visualization code based on users' natural language descriptions~\cite{maddigan2023chat2vis, noever2023numeracy,Dibia2023LIDA}.
For example,
Maddigan et al.~\cite{maddigan2023chat2vis} developed Chat2VIS,
an LLM-based system that generates visualizations based on natural language queries, such as \emph{``Show debt and earnings for Public and Private colleges.''}
\cmo{VizGPT~\cite{VizGpt} and ChartGPT~\cite{ChartGPT} both enable users to create contextual data visualizations through a chat interface from tabular datasets.
Daigr.am~\cite{Daigram} is a GPT plugin that allow users to create visualizations in ChatGPT.
}
However, to the best of our knowledge, 
no previous work has explored the capabilities of LLMs in diverse visualization tasks beyond generating visualization code. 
In this study, 
we conduct a preliminary investigation into the potential of GPT in a broader range of visualization tasks, hoping to inspire future research in this area.

\section{GPT for CS171 - a Data Visualization Course}

This section will first introduce Harvard's undergraduate visualization course CS171 and then outline the settings of our experiments.

\subsection{CS171 at Harvard}

CS171 is an introductory data visualization course at Harvard University. 
The course covers key design principles, design processes, and web-based programming skills. 
Student's grades are determined based on their participation in class activities (10\%) and their performance on the following tasks:

    \para{Quizzes (25\%)} are a set of objective questions, including single-choice, multiple-choice, and matching questions.
    These questions cover a range of visualization knowledge, including design principles, web programming, and practical applications. 
    CS171 contains 91 quiz questions that can all be graded automatically.


    \para{Homework (30\%)} are subjective questions requiring students to write essays and (or) design, implement, and evaluate interactive visualizations. Each assignment includes background information, task descriptions, datasets,  code templates, and instructions. 
    TFs grade assignments using rubrics. 
    CS171 includes nine assignments spanning nine weeks.

    \para{Final Project (35\%)} Students work in teams of three to design and implement an interactive visualization website on a topic of their choice.
    The final project is graded based on the quality of the visualizations, 
    the effectiveness of the data storytelling, 
    and the creativity and innovation of the design.

\subsection{Evaluating GPT on CS171}
We evaluated two GPT models, GPT-3.5 and GPT-4, on the course components. 
To accurately assess the inherent capabilities of GPT models, 
we conducted a zero-shot evaluation without employing complex prompting strategies or domain-specific fine-tuning. 
Specifically, we designed our experiments as follows (All prompts and results can be found in the supplemental materials):

\para{Quizzes.}
The prompt template for quizzes is shown below. 
The \texttt{question type} indicates whether it is a single-choice, multi-choice, or matching question.
The \texttt{format requirements} is decided based on the question type.
We repeat each question 10 times and calculate the average grade.

\begin{coloredquotation}
    \textbf{Prompt template for quiz:}\\
    The following is a quiz question, please give your answer in the following format \{ \texttt{format requirements} \}.\\
    \{ \texttt{question type} \} \\
    \{ \texttt{quiz question} \}
\end{coloredquotation}

\para{Homework.}
The prompt template for homework is shown below. 
To accommodate the token length limitation, GPT receives each step sequentially from the step-by-step instructions.
When necessary, the content of file templates referenced in the instructions is provided.

\begin{coloredquotation}
    \textbf{Prompt template for homework:}\\
    \{  \texttt{background of the homework} \} \\
    \{ \texttt{content of the related template file} \} \\
    \{  \texttt{one step of the instructions} \} \\
    Please output the changes you made to \{\texttt{name of the template file}  \}
\end{coloredquotation}

We added the code generated by GPT into template files and executed them in a browser. 
If an error arose, we re-generated the response; otherwise, we moved forward to the next instruction.
After 30 failed attempts, we ceased re-tries for GPT.

For the grading process, three CS171 TFs, who are also coauthors, independently assessed the assignments using a rubric that has been employed in the course for many years. 
To reduce bias, assignments from GPT were mixed with past student submissions, and the three TFs were unaware of their origin. 
Each assignment was graded by two TFs, and the scores were then averaged.
After grading an assignment, a TF assessed whether the assignment was generated by a human or an AI using a 5-point Likert scale (1 indicates ``definitely human'' and 5 indicates ``definitely AI''). 
TFs also offered free-form comments that addressed various aspects of the assignment, including the structure of the code, the quality of the documentation, and the creativity demonstrated in the work.

\para{Final Project.}
We did not directly evaluate GPT on the final project due to its team-based nature and its reliance on peer evaluation.
Instead, we evaluated GPT at various stages of a visualization project, 
including data collection, data cleanup, visualization design, visual exploration, and insight communication.
Below is one example we used to prompt GPT to explore a chart interactively.
We present examples and share insights into GPT's capabilities that we have discovered through our open-ended experiments.


\begin{coloredquotation}

    \textbf{Prompt template for open-ended explorations:}\\
    This is what you see in a data visualization: \{ \texttt{svg content} \}.\\
    And the following is its JavaScript code: \{ \texttt{JavaScript code} \}.\\
    Now we want to find some insights from the interactive visualization. Please tell me all the actions you want to take in this exploratory process, and show me the JavaScript code to dispatch events and interact with the chart.

\end{coloredquotation}

\section{Results}

\setlength{\tabcolsep}{1pt}
\begin{table}[h]
    \centering
    \small
    \caption{GPTs' scores on the Quizzes and Homework, with a maximum score of 100\% for each column. Each homework is hyperlinked for reference.}
    \label{table:scenes}
    \begin{tabular}{l|
    >{\centering\arraybackslash}p{11mm}|
    >{\centering\arraybackslash}p{4mm}
    >{\centering\arraybackslash}p{4mm}
    >{\centering\arraybackslash}p{4mm}
    >{\centering\arraybackslash}p{4mm}
    >{\centering\arraybackslash}p{4mm}
    >{\centering\arraybackslash}p{4mm}
    >{\centering\arraybackslash}p{4mm}
    >{\centering\arraybackslash}p{4mm}
    >{\centering\arraybackslash}p{4mm}|
    >{\centering\arraybackslash}p{14mm} 
}
    \toprule

     & \textbf{Quizzes} & \multicolumn{9}{c|}{\textbf{Homework}} & \textbf{Weight Sum} \\ 
     &   & \MYhref{https://www.cs171.org/Homework_instructions/week-01/week_01_hw.html}{W1} & 
     \MYhref{https://www.cs171.org/Homework_instructions/week-02/hw/HW2.html}{W2} & 
     \MYhref{https://www.cs171.org/Homework_instructions/week-03/hw/week-03_hw.html}{W3} & 
     \MYhref{https://www.cs171.org/Homework_instructions/week-04/hw/week-04_hw.html}{W4} &
     \MYhref{https://www.cs171.org/Homework_instructions/week-05/hw/week_05_hw.html}{W5} &
     \MYhref{https://www.cs171.org/Homework_instructions/week-06/hw/week-06_hw.html}{W6} & 
     \MYhref{https://www.cs171.org/Homework_instructions/week-07-design/week-07_design.html}{W7} & 
     \MYhref{https://www.cs171.org/Homework_instructions/week-08/hw/week-08_hw.html}{W8} & 
     \MYhref{https://www.cs171.org/Homework_instructions/week-09/hw/week-09_hw.html}{W9} & \\ \midrule
    \textbf{GPT-3.5} & 68.8 & 80 & 68 & 98 & 86 & 100 & 91 & 53 & 85 & 9 &   72.3 \\
    \textbf{GPT-4} & 86.4  & 80 & 86 & 70 & 91 & 86 & 66 & 71 & 82 & 73 & 
    \textbf{82.7} \\ 
    \bottomrule
    \end{tabular}
\end{table}

\subsection{Quiz}

Both GPT-3.5 and GPT-4 completed most of the quizzes.
GPT-3.5 scored 68.8\% while GPT-4 scored 86.4\%.
These results overall indicate that GPT has a wide range of visualization knowledge, 
including design principles, web programming, and practical applications.
Below we discuss the failure cases in finishing the quizzes:

\begin{itemize}[leftmargin=*]
\item \textbf{Cannot read figures.}
Since the GPTs we utilized were incapable of taking images as input, 
they did not perform well on questions that required extracting information from images. 
Nevertheless, we are optimistic that a multi-modal GPT (such as the full version of GPT-4) will excel in these types of questions in the near future.

\item \textbf{Bad at matching questions.}
We discovered that GPT did not excel in matching questions.
We suspect that the text descriptions provided in these questions are too ambiguous for GPT to comprehend and solve. 
However, humans can discern the intention of the questions from the user interfaces (i.e., the drop-down list). 
We believe that if a future version of GPT is capable of reading user interfaces, its performance in these tasks will be enhanced.

\end{itemize}

\subsection{Homework}

Both GPT-3.5 and GPT-4 were able to complete most of the homework assignments, with GPT-3.5 achieving a score of 75.2\% and GPT-4 achieving a score of 79.7\%. In the following section, we will discuss our observations and the feedback we received from the TFs.

\subsubsection{Observations in Using GPT}
GPT clearly showed the capability to understand instructions and complete assignments. 
Through these assignments, 
GPT demonstrated knowledge of visualization theory (e.g., exploratory vs. explanatory visualizations, Gestalt principles), 
visualization design (e.g., design critiques),
web programming skills (e.g., HTML/CSS/JavaScript),
data transformations,
and visualization-dedicated libraries (e.g., D3).
GPT-4 outperformed GPT-3.5 in many ways.
For example,
while GPT-3.5 typically needed dozens of attempts to develop one error-free solution, 
GPT-4 could come up with the correct solution in less than 10 retries.

Below we summarize the challenges we faced when using GPT to finish CS171 homework assignments.
\begin{enumerate}[leftmargin=*, itemsep=0em, label=\texttt{C}\arabic*.]
    \item \textbf{Associating different code files.}
    For example, 
    when tasked to finish a programming task,
    GPT could refer to D3 version 7 in the HTML file but use the APIs from D3 version 6 in the JavaScript (JS) file.
    This implies that GPT cannot reason about the relationship between the code files it generates.
    However, when the instructions explicitly mention the version of the libraries used, GPT can correctly generate the code.
    
    \item \textbf{Outdated Knowledge.}
    Since GPT is trained on data collected before September 2021, it may generate outdated information.
    For example, Homework 1 asks for examples of good and bad visualizations and explanations of why they are good or bad. 
    While the GPT generated reasonable explanations,
    the URL links to the visualizations were expired.
    
    \item \textbf{Hallucination.} 
    The GPT versions we used do not support images as input. 
    Thus, for assignments that contained images in their instructions, we simply provided a URL of the image.
    For example, in the assignment Week 2, GPT is asked to criticize a visualization of the top 10 salaries at Google. 
    Instead of acknowledging that it cannot read the image, GPT will ``imagine'' an unexist visualization and discuss the pros and cons of it.
    This is the so-called ``hallucination'' phenomenon of GPT.
    
    \item \textbf{Ethical Constraints.} 
    Moderation layers have been introduced to GPT to prevent content that violates OpenAI's usage policies. 
    From time to time, GPT responded that it cannot generate the code or answers for us since it is unethical to cheat in doing homework.
    Nevertheless, this can simply be resolved by re-trying a few more times.
    
    \item{\textbf{Token Length Limitation.}}
    GPT has a limit on the token length (i.e., 4k for GPT-3.5, and 8k GPT-4). Thus, we sometimes could not input all the instructions, datasets, and code templates into GPT at once and had to break them into small chunks.
    However, the tasks in Week 9 involved too many lines of code and could not meet the token limit, even when breaking the tasks into smaller pieces.
    Thus, GPT-3.5 could not finish the homework for Week 9.
    
\end{enumerate}

\subsubsection{Feedback from Grading GPT}
Overall, TFs were impressed by the quality of the GPT-generated homework assignments, which is also reflected by the grades (GPT-3.5 scored 75.2\% and GPT-4 scored 79.7\%).

When categorizing the homework as completed by either a human or GPT,
among the total of 54 graded assessments (3 TFs $\times$ 9 HWs $\times$ 2 submissions), 
TFs identified 38 correctly, were incorrect on 8, and were unsure on 8.
Such a high accuracy contrasts with the findings of previous research that evaluated humans' ability to distinguish GPT-generated texts from human-written texts in general tasks~\cite{jakesch2023human, clark2021all}, where humans showed similar accuracy as random guessing.
We hypothesize that this discrepancy in accuracy may be attributed to the distinctions between professional and non-professional contexts and the inherent differences between natural language text and programming code.
At the same time, GPT demonstrated a notable capacity to mislead the TFs to some degree.
When TFs categorized the homework as \textit{``maybe human''}, the accuracy was only 33.3\% (4 correct identifications out of 12). In other words, on 8 out of the 12 occasions when TFs classified the homework assignments as \textit{``maybe human''}, the assignments were, in reality, generated by GPT.


\begin{figure*}[b]
    \centering
    \vspace{-2mm}
    \includegraphics[width=\textwidth]{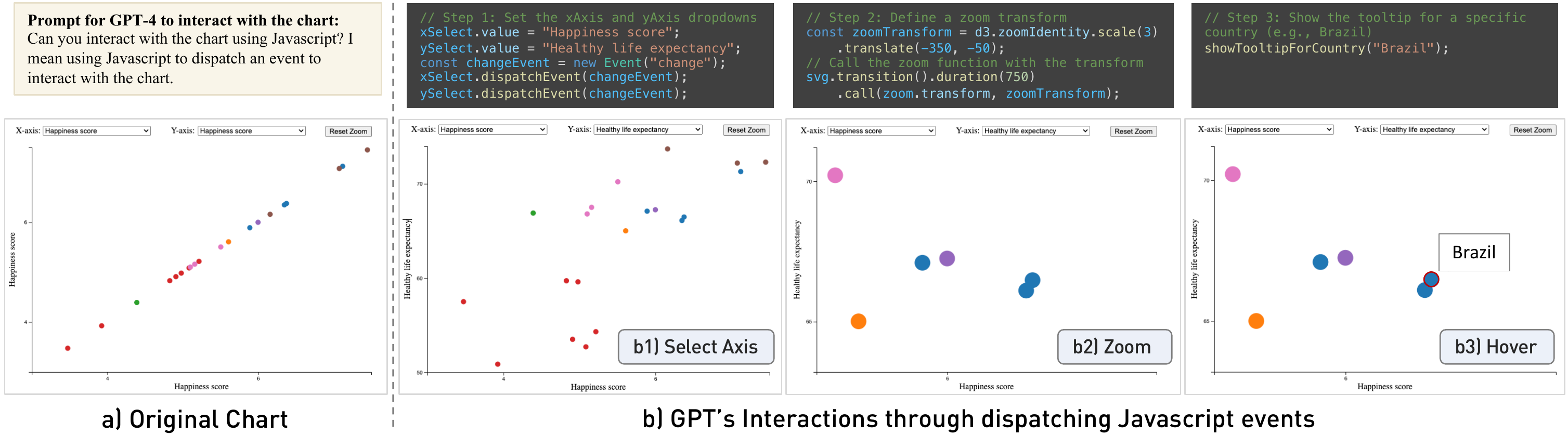}
    \vspace{-4mm}
    \caption{GPT-4 can interactively explore a visualization by dispatching JS events.}
    \label{fig:interaction_examples}
\end{figure*}

The three TFs reported the heuristics they used to distinguish between assignments submitted by students and those generated by GPT. 
While these heuristics reveal valuable insights about GPT-generated assignments, they were not consistently true.
\begin{itemize}[leftmargin=*]
    \item \textbf{GPT lacks creativity [true].} TFs were able to identify student submissions based on the belief that \textit{``this level of creativity is not exhibited by GPT models''} (TF3) and that GPT-generated homework tended to be \textit{``too generic (and) could be applied to anything''}. Student submissions exhibit creativity through diverse elements such as unique color selection, attractive animation effects, thoughtful variable naming, and a cohesive aesthetic.
    
    \item \textbf{GPT shows atypical usage of programming libraries [true].} TFs tended to believe the assignments were generated by GPT if they observed atypical usage of programming libraries. For example, TF2 successfully identified a GPT-generated homework, because \textit{``the mix of D3 and vanilla js in the main.js is somewhat odd''}.
    
    \item \textbf{GPT fails to read the visualization accurately [true].} 
    TFs were able to identify submissions generated by GPT due to the model's inability to accurately interpret the content depicted in visualizations. For instance, GPT-generated textual descriptions might inaccurately refer to bar charts, even when there are no bar charts present in the visualization. 
    GPT also generated visualizations in which the content was improperly positioned, e.g., \textit{``the map is half off the screen''} (TF2). 
    This feedback echos \texttt{C}3 in Sec.4.2.1.

    \item \textbf{GPT does not write informal content [false].} 
    TFs generally held the belief that GPT does not produce informal or disorganized content. 
    For example, TF1 commented \textit{``some comments are informal enough that they seem student written''}. As a result, they mistakenly categorized some submissions as \textit{``maybe human''} when evaluating them. 
\end{itemize}

It is noteworthy that these observations should be interpreted with caution due to the limited sample size, the selection of only high-quality student assignments, and the minimal prompt engineering employed in utilizing GPT.

\subsection{Final Project}
We conducted an open-ended study to explore GPT-4's capabilities in completing typical visualization tasks required to finish the final project of CS171. 
Below we summarize our key findings.



\para{Collecting and Cleaning Data.} 
GPT-4 can assist with data collection and cleanup. 
It suggested websites and specific datasets to download. 
Additionally, GPT-4 successfully generated Python code (using libraries like pandas and numpy) to clean the datasets, merge them, handle missing or inconsistent data, derive metrics, and save the cleaned data into new files.

\para{Designing Exploratory Visualizations based on Datasets and User Requirements.}
We tested GPT's capability of designing exploratory visualizations based on the provided data and user requirements.
We provided a toy dataset from the World Happiness Report 2020 and asked GPT-4 to create an interactive visualization that allows users
to browse, filter, and compare different data.
GPT-4 created an interactive scatter plot using D3.js, 
which allows users to select the X and Y axes using dropdown menus.
When the user hovers over a data point, a tooltip displays the country, region, and the selected axes' values.
GPT-4 can also improve the visualizations based on user feedback. 
For instance, when we expressed the need to examine specific countries more closely within a graph, GPT-4 implemented a zooming interaction.


\para{Reading Visualizations in SVG and JS.}
While the GPT-4 model we used does not support images as input,
we found that it was capable of reading visualizations in SVG format.
To test this capability, we provided GPT with a set of scatter plots in SVG format and asked it to describe its observations.
GPT-4 was able to correctly describe the chart type,
mark shapes (e.g., circle),
encodings (e.g., x- and y-axis, and color of the circles),
and the range of the axes. 
In addition to the SVG, we further provided GPT-4 with the JS code of the scatter plot and asked it to describe the available interactions that can be performed on the visualization.
It successfully identified the available interactions, such as switching which data attributes are mapped onto an axis through interactions with the drop-down menu and zooming in on the scatter plot.

\para{Interacting with Visualizations through Dispatching JS Events.}
We also asked GPT-4 to interact with the chart by dispatching JS events.
It successfully generated code to update axes by selecting the dropdown menu,
reset the zoom, 
and mimic a hover event to show the tooltip of a country (\autoref{fig:interaction_examples}).
This means that GPT can interact with visualizations by triggering corresponding events in the underlying JS code. 
With the ability to interact with visualizations, GPT could potentially support a wider range of tasks that require user engagement and exploration of data. 

\para{Discovering Insights by Interactive Exploration.}
For insight discovery, 
we asked GPT-4 to generate codes to interact with a scatter plot and feed the resulting SVG file to it.
It not only can interact with the chart and observe the correlation, trend, clusters,
and outliers from the scatter plot, 
but also contextualize these insights with its extensive background knowledge.
For example, it identified that Gabon ``\emph{has a relatively high Logged GDP per capital value but a lower Healthy life expectancy compared to other countries with similar GDP per capital values}.''
It further explained this insight by that Gabon is a country known for its rich natural resources, particularly oil and timber, but has limited access to healthcare.


\begin{figure}[h]
    \centering
    \vspace{-3mm}
    \includegraphics[width=0.8\columnwidth]{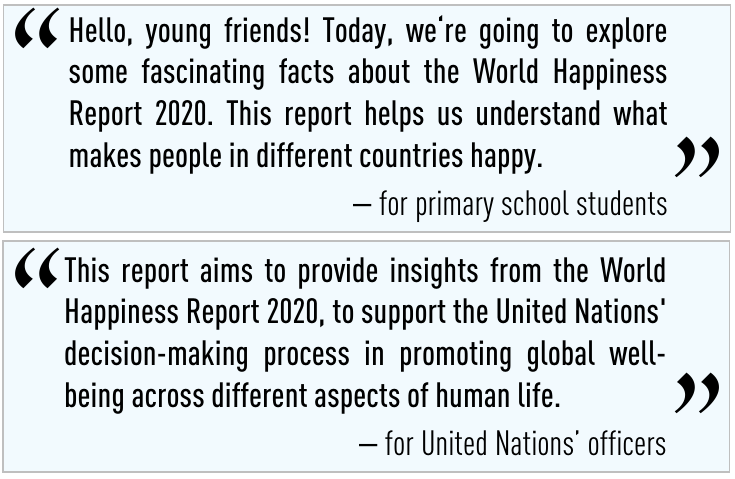}
    \vspace{-3mm}
    \caption{GPT-4 presents insights with different tones for different target readers: conversational for primary school students (Top) and formal for UN officers (Bottom).}
    \label{fig:tone_examples}
\end{figure}

\para{Interpreting and Presenting Data Insights for Different Targeted Users.}
To test GPT-4's ability to present insights for various targeted audiences, we provided a set of scatter plots in SVG format and asked it to generate reports for the government of African countries, the UN, and primary school students (\cmo{\autoref{fig:tone_examples}}). 
GPT-4 selected the relevant insights from the charts and presented them in an appropriate tone for each target reader. 
This capability to generate tailored reports could have practical applications.




\begin{figure}[h]
    \centering
    \vspace{-2mm}
    \includegraphics[width=0.99\columnwidth]{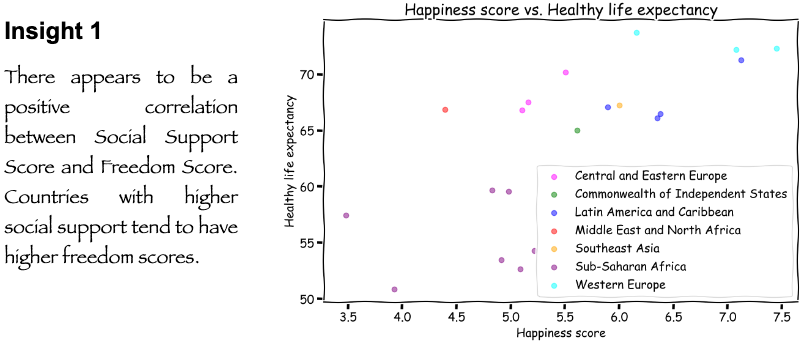}
    \vspace{-2mm}
    \caption{GPT-4 can communicate insights using scrolly-telling style webpage.}
    \label{fig:story_examples}
\end{figure}

\para{Telling a Data Story.}
GPT-4 can generate narrative visualizations when provided with a dataset and a list of insights (\cmo{\autoref{fig:story_examples}}). 
By default, GPT-4 may opt for Python libraries and annotated scatter plots to create these visualizations. 
However, when specifically requested to implement scrolly-telling, 
GPT-4 can generate an HTML-based scrolly-telling visualization using D3.js and Scrollama.
Other more complicated narrative formats, such as data movies, present a challenge for an end-to-end generation.
However, GPT acknowledges this limitation and offers an alternative solution, responding ``\emph{while it's not possible to create a full data video directly within this text response ... (I) can guide you through the process of creating one.}''

\section{Discussion and Conclusion}

Although our study has provided valuable insights into GPT's performance in visualization tasks, it is important to acknowledge its limitations, including the small sample size, limited prompt engineering, 
comparison only with high-quality student assignments, 
and the specificity of the GPT models used. 
Nonetheless, we believe that this preliminary study can stimulate further research and serve as a foundation for the visualization community. 
Due to page limitations, we cannot fully expand our discussion here. 
Thus, we outline below the main points that warrant future research.




\para{Strength and Limitations of GPT in Visualization.}
Perhaps the biggest strength of GPT is its ability to \emph{reason} about users' natural language input and generate high-quality textual output. 
Furthermore, its potential in visualization remains untapped due to the limited prompt engineering and fine-tuning we used. 
However, 
while limitations such as token length limit and incapability to process image input can be fixed in the near future, others can limit GPT's usage in visualization tasks.
For example, 
its inability to associate multiple code files can limit its usage in large-scale visual analytic systems; the unstable output of GPT can require multiple attempts to generate satisfactory results.
Although GPT has shown promise in visualization tasks, more work is needed to identify, evaluate, and overcome its limitations and fully realize its potential in the field.





\para{Facilitating Broader Visualization Tasks with LLMs} presents a promising future direction in the field.
Through our investigations, 
we found that GPT possesses rich knowledge of visualization design, and can be used to design exploratory visualizations, 
read visualizations in SVG format, 
interact with visualizations through dispatching JS events, 
and even create fancy narrative visualizations for general audiences.
These findings highlight the vast capabilities of LLMs and their potential to automate complex visualization tasks. 
In this sense, LLMs can be viewed as ``visualization experts'' that can provide valuable assistance to novices in visual data tasks. 
The potential of LLMs to enhance a broad range of visualization tasks presents exciting opportunities for future research.

\para{\cmo{Ethical Constraints of GPT Models.}}
\cmo{
While our research primarily focused on the capabilities of the GPT model in data visualization, we acknowledge that GPT and similar models have garnered concern over their inception and deployment, particularly around issues of equity and inclusion. As these models learn from vast amounts of data on the internet, they may inadvertently perpetuate biases present in these datasets. 
This raises valid concerns about how these models might influence users, 
especially if they reinforce existing societal biases. 
Such biases could skew students' perspectives if unchecked in educational settings. 
It is crucial to approach these tools with a discerning mindset, ensuring that the visualizations they inform are free from unintentional prejudice and promote inclusivity. Future work in this area should prioritize examining the outputs of such models through an ethical lens and implementing measures to correct any identified biases.
}

\para{Redesigning Visualization Education} is an important task for the community.
Our experiments show that students in a visualization class can easily score well without truly learning the covered material, highlighting the need to rethink and redesign the way visualization is taught.
\cmo{One approach is to place emphasis on areas where human judgment and nuanced interpretation are irreplaceable.}
\cmo{For example, instructors could include tasks that are hard or impossible for GPT to perform alone, such as open-ended questions that require skills like sketching or unique information provided in class, or incorporate more group projects that promote teamwork and collaboration.}
On the other hand, 
\cmo{instructors could consider integrating these AI tools into the curriculum itself, allowing students to learn alongside and from these models, while also critically assessing their outputs.}
For example, proper prompts from students can inspire GPT to be more creative; 
teaching students how to use GPT to solve visualization problems could provide a human/AI collaborative and personalized approach to learning.
\cmo{For CS171, we aim to incorporate GPT as a supplementary tool for students, guiding them to harness its capabilities for visualization challenges.}
Overall, redesigning visualization courses is essential to better evaluate students' learning and promote their engagement with the material.




\bibliographystyle{abbrv-doi}

\bibliography{template}
\end{document}